\documentclass[prc,preprint,showpacs,doublespaced,article]{revtex4}
\usepackage{graphicx}
\usepackage{epstopdf}
\usepackage{color}
\usepackage{amsmath}
\usepackage{multirow}
\begin{document}
\title{{Binary neutron star mergers within kaon condensation: GW170817}}

\author{{ \bf Z. Sharifi\textsuperscript{1}\footnote{E-mail: sharifi@znu.ac.ir}, \bf M. Bigdeli\textsuperscript{1}\footnote{E-mail: m\underline\ \ bigdeli@znu.ac.ir}, \bf D. Alvarez-Castillo\textsuperscript{2,3}, \bf E. Nasiri\textsuperscript{1}}}
\affiliation{\textsuperscript{1}Department of Physics, University of Zanjan, P.O. Box
45371-38791 Zanjan, Iran}
\affiliation{\textsuperscript{2} {H. Niewod\'{n}iczanski Institute of Nuclear Physics,
Radzikowskiego 152, 31-342 Krak\'{o}w, Poland}}
\affiliation{\textsuperscript{3} Bogoliubov Laboratory for Theoretical Physics,
Joint Institute for Nuclear Research, 6 Joliot-Curie Street, 141980 Dubna, Russia}

\begin{abstract}
The equation of state (EoS) of neutron star (NS) matter is investigated considering kaon condensation. Moreover, the tidal parameters related to the  components of binary neutron star mergers are determined and compared to the constraints of GW170817 imposed on these quantities. In this study, we employ the lowest-order constrained variational (LOCV) approach and utilize {AV6$'$, AV8$'$, and AV18 potentials} accompanied by the three nucleon interaction (TNI)  {in order to consider the nucleon-nucleon interaction}. It is known that the existence of kaons in the core of NSs softens the EoS  {and thus} lowers the value of maximum mass which is expected to be greater than 2 M$_\odot$. Our results demonstrate that considering kaon condensation with a  strangeness value, a$_{3}$m$_s$=-134 MeV satisfies the maximum mass constraint except for AV8$'$+TNI potential. However, the calculation of dimensionless tidal deformability shows that with the decrease of strangeness value, the neutron star gets less deformed.


\textbf{Kewords:} Kaon condensation, binary neutron star mergers, GW170817, LOCV method

\end{abstract}

\maketitle
\section{INTRODUCTION}
GW170817 as the first gravitational wave detection resulting from a binary neutron star merger and its electromagnetic counterpart has provided valuable information concerning dense matter~\cite{abott}. Other recent multi-messenger observations, such as GW190814~\cite{190814} and GW190425~\cite{190425} have also set some valuable constraints on the structure and tidal deformability of compact stars. Rezzolla \textit{et al.} set limits for the maximum mass of non-rotating stellar configurations using the observation of the GW170817 event, modeling of GRB 170817A, and the quasi-universal relations~\cite{Rezzollaa}. Furthermore, Bauswein \textit{et al.} investigated that the radius of the stellar structure of 1.6 M$_{\odot}$ non-rotating NSs must be larger than 10.68$_{-0.04}^{+0.15}$ km, which is obtained by the binary mass measurement of GW170817 and the assumption of delayed collapse in this event~\cite{Bauswein}. The maximum radius of a 1.4 M$_{\odot}$ NS is also obtained to be 13.6 km~\cite{Annala}. According to the most recent detection GW190814, Most \textit{et al.} estimated a lower boundary on the maximum mass of non-rotating neutron stars~\cite{Mostt}. {Furthermore, the new NICER mass-radius estimation of PSR J0740+6620 sets constraints on the EoS \cite{Raaijmakers}. The 95\% credible ranges of the radius of 1.4 M$_{\odot}$ NS are found to be 12.33$_{-0.81}^{+0.76}$ km for a piecewise polytropic (PP) model and 12.18$_{-0.79}^{+0.56}$ km for a model based on the speed of sound in a NS (CS) \cite{Raaijmakers}.}

The EoS of NS's core has a major impact on the value of maximum mass. Modeling NSs at high densities (about ten times normal nuclear density, $\rho_{0}=0.16 fm^{-3}$) is of great importance in which the existence of exotic matter, such as hyperons, meson condensate, and quark matter is probable~\cite{1.}. {Thermal effects are small when the maximum neutron star mass is considered in all these cases. When matter contains negatively charged particles, namely a kaon condensate, neutrino trapping leads to an increase in the maximum mass~\cite{1.}.} It is well-known that the appearance of hyperons in the core of NSs results in a considerable decrease of their maximum mass~\cite{2., 3.}. In addition, the possibility that kaon condensation occurs in the dense matter has also been investigated by many authors~\cite{4.,5.,6.}, which was first proposed by Kaplan and Nelson~\cite{kaplan}. The existence of a kaon condensate in NS matter softens the EoS, which lowers the value of maximum mass. It is to be noted that attractive interactions between kaons and nucleons (the so-called sigma term, $\Sigma^{KN}$) is the main source of kaon condensation~\cite{7.}.

 {The kinetics of kaon condensation was taken into account employing rate equations including three weak reactions: the thermal kaon process, the kaon-induced Urca process, and the modified Urca process \cite{Muto2000}. It was found that throughout the equilibration process, the thermal kaon process plays an important role in comparison to other weak reactions. The neutrino mean free paths related to protoneutron stars were obtained in the presence of kaon condensates considering both non-degenerate and degenerate neutrino cases \cite{Muto2003}. It was concluded that the kaon-induced neutrino absorption process does not have a key role in the thermal and dynamical evolution relevant to the protoneutron stars.} Very recently, Thapa and Sinha have investigated the possibility of (anti)kaon condensation in $\beta$-stable matter using non-linear and density dependent covariant density functional model~\cite{9.}, which demonstrates consistent results with recent observations. In Ref.~\cite{6.}, nuclear equations of state were calculated by Skyrme force model parameters, and the possibility of kaon condensation in the core of neutron stars was also examined. The authors in Refs.~\cite{4., 10.} utilized the Brueckner-Hartree-Fock (BHF) approximation with realistic AV18 potential and three-body forces to obtain the EoS of kaon condensed phase. Considering the kaon condensed NS matter, the LOCV approach was applied for the nucleon-nucleon energy density and the nuclear symmetry energy using the AV14 and AV18 potentials by Bigdeli \textit{et al.}~\cite{BZN}. In this manuscript, we also employ the LOCV method in order to calculate the nucleon-nucleon interaction with the aid of two-body potentials such as AV6$'$, AV8$'$, and AV18 together with the three nucleon interactions. For the kaon condensation phase, the chiral lagrangian introduced by Kaplan and Nelson is applied to calculate the nucleon-kaon interaction and therefore its energy density. It turns out that by employing TNI the results obtained for the threshold densities remain approximately constant for the different values of strangeness.
In this work, we also intend to explore the effect of kaon condensation on our previously obtained results for the mass-radius relation of NSs~\cite{Asadi2018} in addition to their tidal deformability in a binary neutron star merger and check the validity of our calculations with the observational constraints.

{The first part of the next section involves the LOCV model for obtaining the energy density of nucleon-nucleon interaction, while in the second part we recall the formalism of nucleon-kaon interaction and present our results for three different EoSs with the candidate strangeness values. Applying the selected EoS models, the results corresponding to the mass-radius relation of neutron stars and the tidal deformability of binary neutron star mergers are presented in Sect.~III, and IV, respectively. Finally, our conclusion and summary are given in Sect.~V.}

\section{Equation of State of Neutron Star Matter
with Kaon Condensation}

We have employed the EoSs of the inner and outer crust of neutron stars introduced in~\cite{Douchin, Baym}. Moreover, the core is assumed here to be composed of leptons, nucleons, and kaons. Therefore, the total energy density of this system can be written as
{\begin{eqnarray}\label{te}
           \varepsilon=\varepsilon_{lep}+\varepsilon_{nucl}+\varepsilon_{KN},
 \end{eqnarray}}
in which $\varepsilon_{lep}$, $\varepsilon_{nucl}$, and $\varepsilon_{KN}$ are the energy density of leptons (muons and electrons), interacting nucleons, and kaon-nucleon interaction, respectively. It is generally well-known that leptons form non-interacting fermi gas and their energy density can easily be calculated~\cite{BZN}. One can obtain the energy density of nucleons as $\varepsilon_{nucl}=\rho(E_{nucl}+m)$. It should be pointed out that $m$ is the nucleon mass, $\rho$ is the total number density of nucleons and, $E_{nucl}$ is the total energy per nucleon which is obtained from the LOCV approach.

In the following, we have briefly discussed the calculation of the energy density of nucleons and kaon-nucleon
interaction, separately.
\subsection{LOCV method}

The total energy per nucleon is written as a function of correlation function $f$ and also its derivatives up to the two-body term as~\cite{Clark}:
{\begin{eqnarray}\label{tener}
           E_{nucl}([f])&=&\frac{1}{A}\frac{\langle\psi|H|\psi\rangle}{\langle\psi|\psi\rangle}
            =\frac{1}{A}\sum_{\tau=n,p}\sum_{k\leq k_{\tau}^{F}}\frac{\hbar^{2}k^{2}}{2m_{\tau}}
           \nonumber \\ &+&\frac{1}{2A}\sum_{ij} \langle ij\left| -{\hbar^{2}}/{2m}[f(12),[\nabla_{12}^{2},f(12)]]
           +f(12)V(12)f(12)\right|ij-ji\rangle.
 \end{eqnarray}}
In this formalism, the trial wave function in the form of $\psi=\cal{F}\phi$ is adopted in which $\phi$ and ${\cal F}={\cal S}\prod _{i>j}f(ij)$ correspond to the slater determinant of $A$ independent nucleons and Jastrow form of $A$-body correlation operator with $\cal S$ being a symmetrizing operator, respectively. It is to be noted that the two-body energy consists of two-body correlation operator $f(12)$~\cite{owen} and two-body potential V(12), which the Argonne family potentials, namely AV6$^{'}$+TNI, AV8$^{'}$+TNI, and AV18+TNI~\cite{Asadi2018} are considered in this study.
Here, TNI refers to the three-nucleon interaction contribution~\cite{TNI}.

\subsection{KAON-NUCLEON INTERACTION}
We apply the model presented by Kaplan and Nelson involving SU3$\times$SU3 chiral lagrangian with the octets of pseudoscalar mesons and baryons for the interaction of kaon in nuclear medium~\cite{kaplan}. Considering only the s-wave kaon-nucleon interactions, this effective lagrangian takes the following form~\cite{Brown, prakash}:
{\begin{eqnarray}\label{kaon}
           L_{KN}&=&\frac{f^{2}}{2}\mu_{K}^{2}sin^{2}\theta-2m_{K}^{2}f^{2}sin^{2}\frac{\theta}{2}+n^{\dag}n(\mu_{K} sin^{2}\frac{\theta}{2}-(2a_{2}+4a_{3})m_{s}sin^{2}\frac{\theta}{2})
           \nonumber \\ &+&p^{\dag}p(2\mu_{K} sin^{2}\frac{\theta}{2}-(2a_{1}+2a_{2}+4a_{3})m_{s}sin^{2}\frac{\theta}{2}),
 \end{eqnarray}}
in which $f$ is the pion decay constant (93 MeV), $\mu_{K}$ is the kaon chemical potential and $\theta=\sqrt{2}v_{K}/f$ is the chiral angle, where $v_{K}$ corresponds to the amplitude of the mean field of $K^{-}$, $\langle K^{-}\rangle=v_{K}e^{-i\mu_{K}t}$, determining the magnitude of the condensate. It should be noticed that $a_{1}$, $a_{2}$, and $a_{3}$ are coefficients related to the interaction terms of the original Kaplan-Nelson lagrangian, leading to the masses splitting of the baryon octet, and $m_{s}$ is the mass of strange quark. The two coupling parameters, $a_{1}m_{s}$ and $a_{2}m_{s}$, are determined from the splitting of strange baryon masses and their values are taken to be -67 and 134 MeV, respectively, as in Ref.~\cite{prakash}. However, there is not enough information about the constant $a_{3}m_{s}$ associated with the poorly-known value of strange content of the proton and the kaon-nucleon sigma term,
{\begin{eqnarray}\label{tener}
           m_{s}\langle \overline{s}s \rangle_{p}=-2(a_{2}+a_{3})m_{s},
\end{eqnarray}}
{\begin{eqnarray}\label{tener}
           \Sigma^{KN}=-\frac{1}{2}(a_{1}+2a_{2}+4a_{3})m_{s}.
 \end{eqnarray}}
Thus, four values are chosen for this parameter in this study as in Ref.~\cite{prakash}: -134, -178, -222, and -310 MeV, corresponding to the strangeness content $\langle \overline{s}s \rangle_{q}$=0, 0.05, 0.1 and 0.2, respectively.
The energy density of kaon-nucleon interaction, $\varepsilon_{KN}$, described by the lagrangian (Eq.~\ref{kaon}) reads,
{\begin{eqnarray}\label{tener}
           \varepsilon_{KN}=f^{2}\frac{\mu_{K}^{2}}{2}sin^{2}\theta+2m_{K}^{2}f^{2}sin^{2}\frac{\theta}{2}+\rho(2a_{1}x_{p}+2a_{2}+4a_{3})m_{s}sin^{2}\frac{\theta}{2}.
 \end{eqnarray}}
By minimization of the total energy density (Eq.~\ref{te}), the ground state parameters such as proton fraction, the kaon chemical potential, etc, can be obtained. Note that the charge neutrality and beta equilibrium conditions in addition to the minimization of energy density provide some constraints on our calculations, {which are given by \cite{7., prakash}
{\begin{eqnarray}\label{tener}
           f^{2}\mu sin^{2}\theta+\rho(1+x_{p})sin^{2}\frac{\theta}{2}+\frac{(\mu^{2}-m_{e}^{2})^\frac{3}{2}}{3\pi^{2}}+\frac{(\mu^{2}-m_{\mu}^{2})^\frac{3}{2}}{3\pi^{2}}-\rho x_{p}=0,
 \end{eqnarray}}
{\begin{eqnarray}\label{tener}
           \mu=\mu_{k}=\mu_{e}=\mu_{\mu}=\mu_{n}-\mu_{p}=\frac{4(1-2x_{p})S(\rho)}{cos^{2}\frac{\theta}{2}}-2a_{1}m_{s}tan^{2}\frac{\theta}{2},
 \end{eqnarray}}
{\begin{eqnarray}\label{tener}
           cos\theta=\frac{1}{f^{2}\mu^{2}}\left(m_{K}^{2}f^{2}+\frac{1}{2}\rho(2a_{1}x_{p}+2a_{2}+4a_{3})m_{s}-\frac{1}{2}{\mu}\rho(1+x_{p})\right).
\end{eqnarray}}
Here $S(\rho)=\frac{1}{8}\left(\frac{\partial^{2}E_{nucl}}{\partial {x_p}^{2}}\right)_{x_p=1/2}$ corresponds to the nuclear symmetry energy.
%
Moreover, the threshold density $\rho_{th}$ at which kaons appear can be calculated through these equations for $\theta=0$. {Table~I presents our results {for {incompressibility coefficient} and symmetry energy at saturation density, its slope \cite{Asadi2018}, and also the threshold density for candidate EoSs in addition to those obtained from the EoS with AV18 potential in \cite{BZN}. Very recently, symmetry energy constraints have been obtained to be $42<L<117$ MeV and $32.5<S_{0}<38.1$ MeV applying ratios of the charged pion spectra \cite{Estee}. One can deduce that our results for L is well agreed with these new data, however those obtained for S$_{0}$ are marginally consistent with them. It is to be noted that AV18 potential leads to the values of S$_{0}$ and L parameters that lie exactly in the reported range.} It is also obvious that as strangeness values decrease, the appearance of kaons takes place earlier for all EoS models. Note that AV6$'$+TNI has the largest threshold density range with the change of a$_{3}$m$_{s}$ from -134 to -310 MeV, while AV18 leads to the smallest range. Furthermore, by comparing the last two models, one can observe that applying TNI does not result in a considerable impact on the threshold densities with different strangeness values. However, the result for a$_{3}$m$_{s}$=-310 MeV remains unchanged considering these two EoS models.}
\begin{table*}
{\caption{{\label{1} {Symmetry energy properties, K$_{0}$, S$_{0}$, and L} \cite{Asadi2018}, together with threshold densities $\rho_{th}$ in the units of MeV and {fm}$^{-3}$, {respectively}. {Threshold densities} are presented for our EoS models considering kaon condensation with different values of a$_{3}$m$_{s}$. The results corresponding to the model with AV18 potential are also presented for comparison \cite{BZN}.}}}
%
\begin{tabular}{cccccccc}
  \hline
   \hline
\multirow{3}{*}{{Model}}  &     \multicolumn{6}{c}{$\rho_{th}$}  \\ \cline{5-8}  & {\multirow{1}{*}{K$_{0}$}} &\multirow{1}{*}{S$_{0}$}  &\multirow{1}{*}{L}
& {a$_3$m$_{s}$=-134 MeV} & {a$_3$m$_{s}$=-178 MeV}  & {a$_3$m$_{s}$=-222 MeV}  & {a$_3$m$_{s}$=-310 MeV}   \\
 \hline
  AV6$'$+TNI&{266}&30.51&43.06&0.889&0.712&0.587&0.429\\
  \hline
AV8$'$+TNI&{253}&29.83&42.52&0.836&0.685&0.574&0.427
\\
   \hline
AV18+TNI&{264}&30.37&48.19&0.814&0.672 &0.565&0.422
\\
  \hline
   \hline
AV18&{301}&36.24&63.07&0.812&- &0.563&0.422
\\
   \hline
   \hline
  \end{tabular}
  \end{table*}

\begin{figure}[!ht]
\centerline{\includegraphics[scale=0.7]{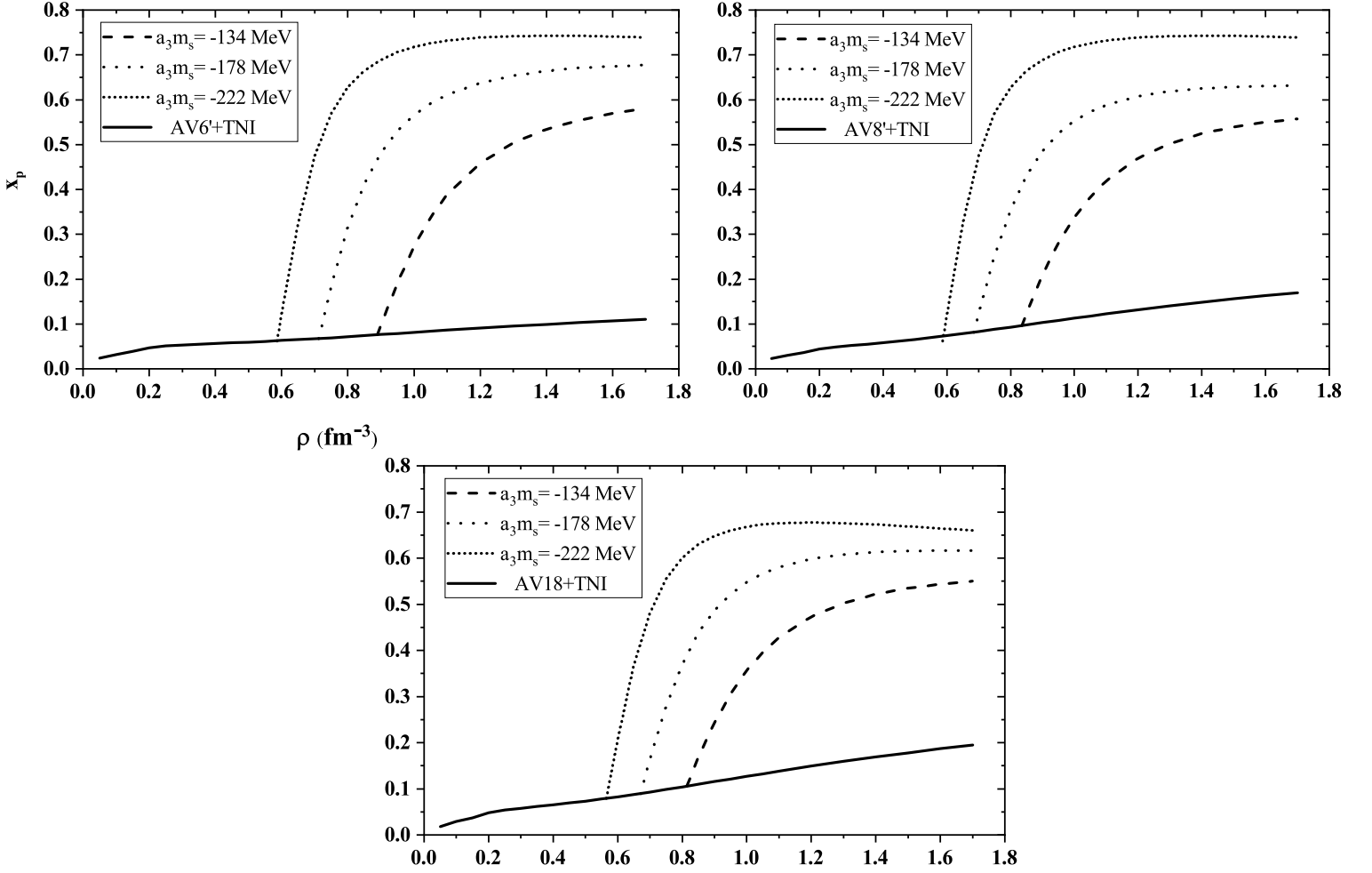}}
\caption{\small {Proton fraction as a function of baryon number density for three potentials of AV6', AV8', and AV18 accompanied by TNI employing a$_{3}$m$_{s}$=-134, -178, and -222 MeV.}}\label{F1}
\end{figure}
\begin{figure}[!ht]
\centerline{\includegraphics[scale=0.6]{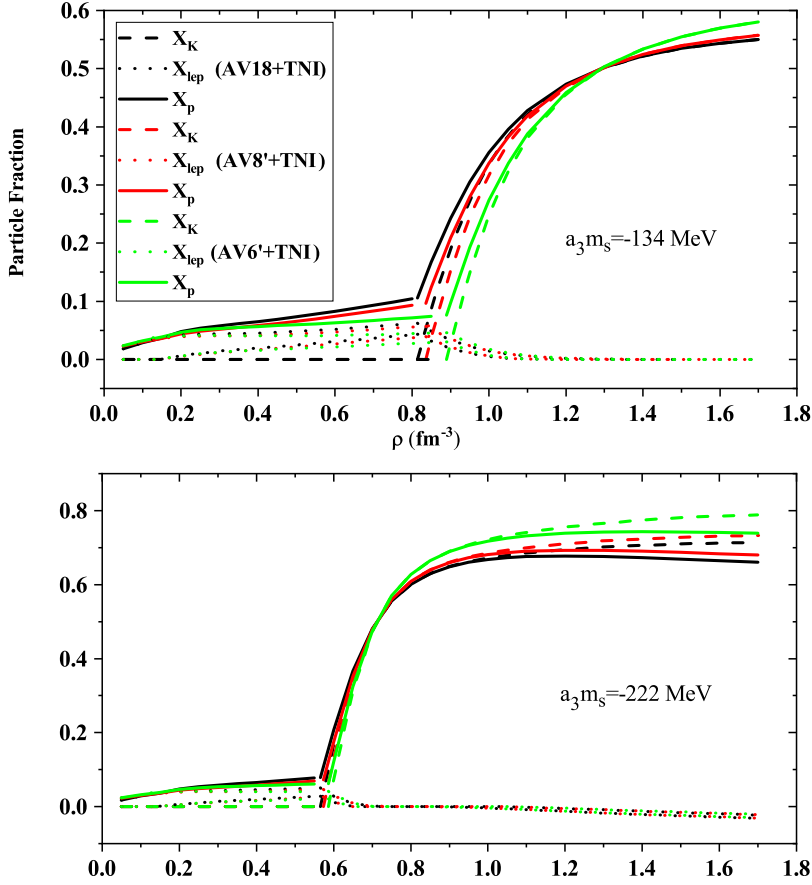}}
\caption{\small {Particle fraction (proton, kaon and lepton fraction) versus baryon number density employing three potentials, AV6'+TNI, AV8'+TNI, and AV18+TNI, with a$_{3}$m$_{s}$=-134 and -222 MeV.}}\label{FFF}
\end{figure}
\begin{figure}[!ht]
\centerline{\includegraphics[scale=0.7]{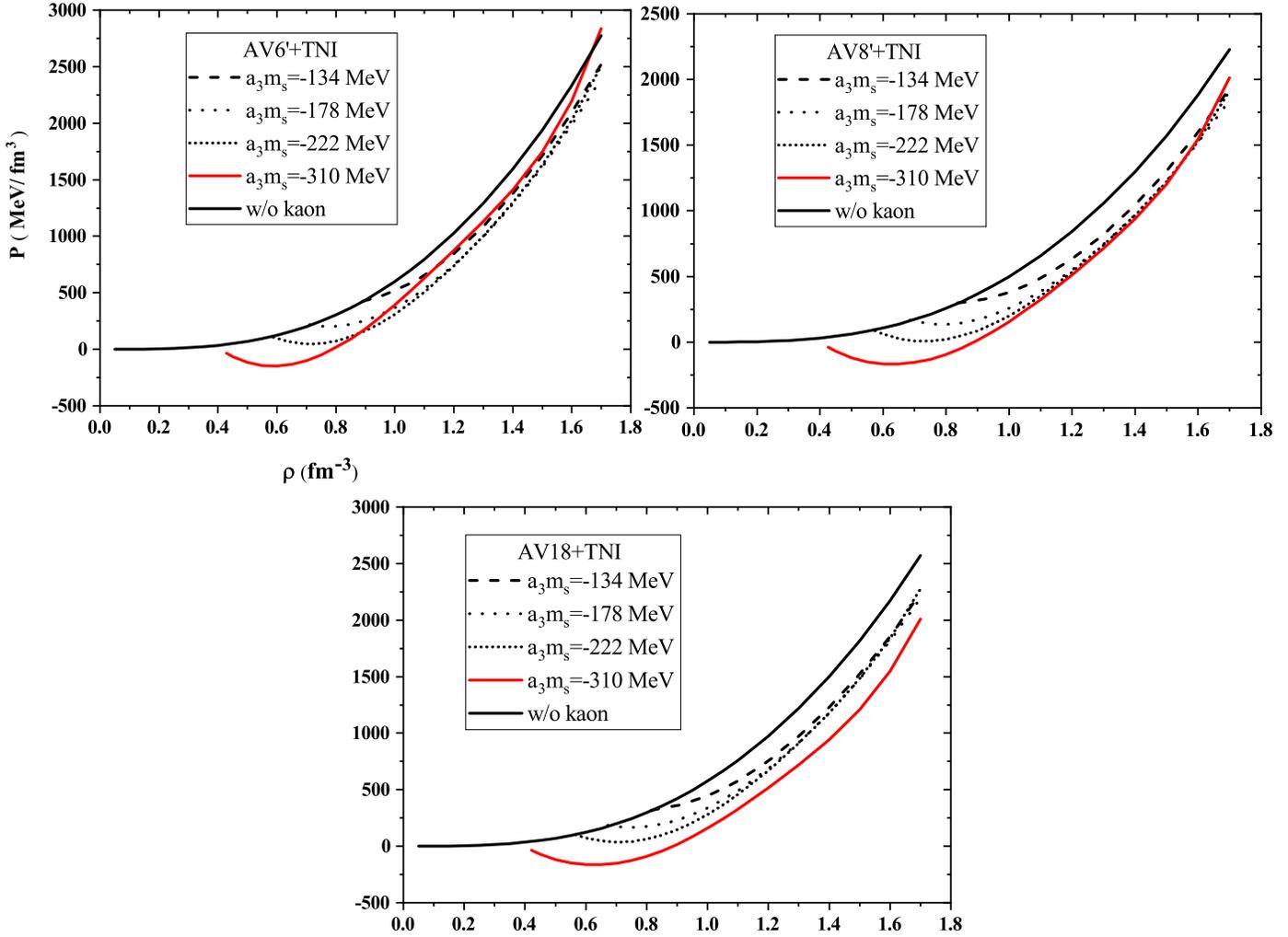}}
\caption{\small {Same as in figure 1, but for the pressure.}}\label{F2}
\end{figure}
\begin{figure}[!ht]
\centerline{\includegraphics[scale=0.7]{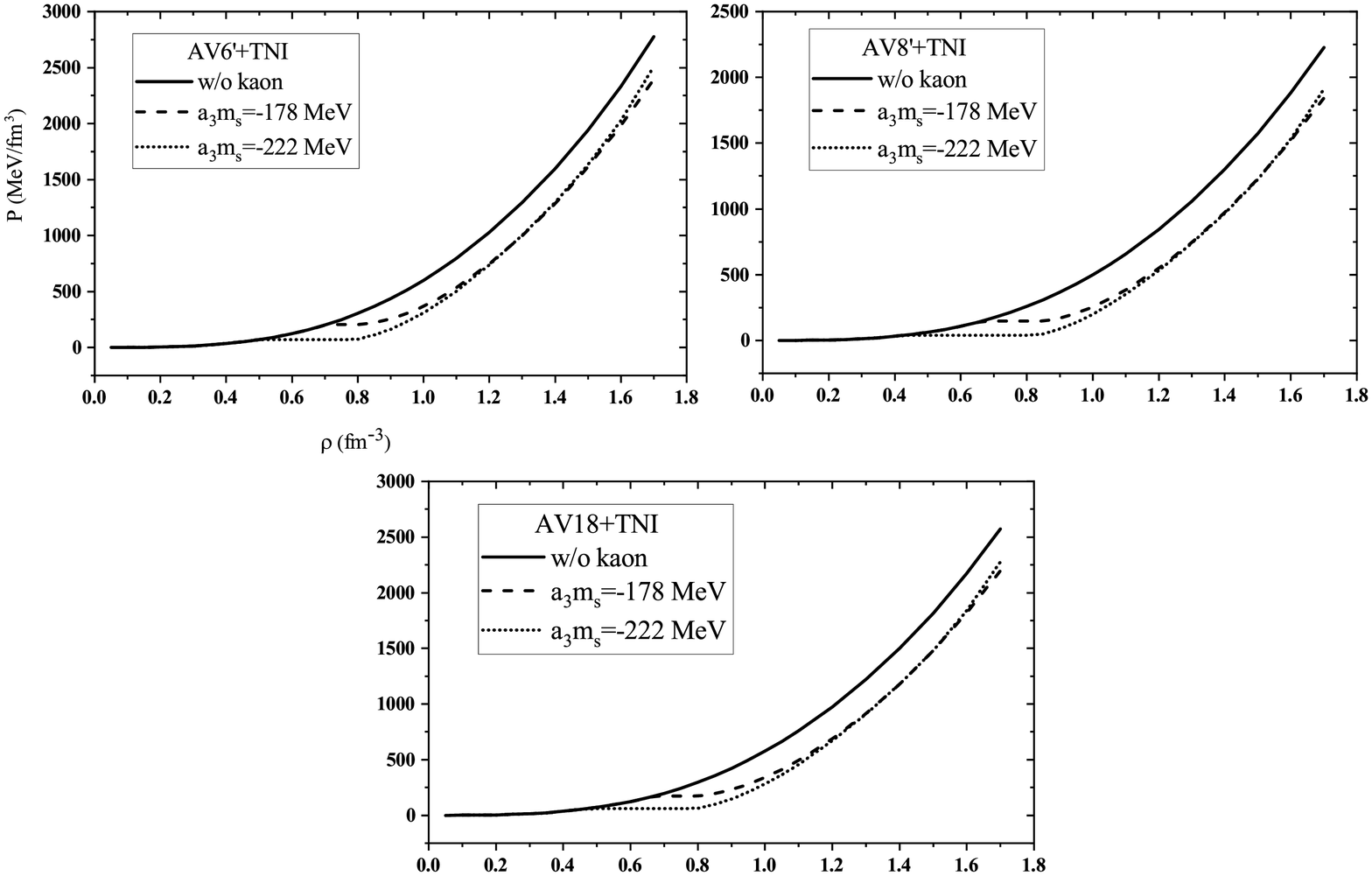}}
\caption{\small {Pressure as a function of baryon number density for the candidate EoSs. Maxwell construction is used for a$_{3}$m$_{s}$=-178, and -222 MeV in order to gain the stability.}}\label{F3}
\end{figure}

We proceed by calculation of the required parameters for obtaining the EoS like {particle} fraction, which is presented in figure~\ref{F1} and {figure~\ref{FFF}} as a function of the baryon number density. Figure~\ref{F1} {(\ref{FFF})} is provided by employing the results of AV6$'$+TNI, AV8$'$+TNI, and AV18+TNI potentials together with different values of a$_{3}$m$_{s}$=-134, -178, and -222 MeV {(a$_{3}$m$_{s}$=-134 and -222 MeV)}. It is apparent from {these figures} that the proton fraction is strongly affected after the appearance of kaons in neutron star matter, where it increases sharply after the threshold density. The main reason for this issue lies in the neutralization of the negative charge of kaons. {It is also seen that lepton fraction decreases after the kaon condensation for all potentials and even becomes negative, which is so clear in the case of a$_{3}$m$_{s}$=-222 MeV. This is due to the appearance of anti-leptons at higher density values, which is more probable as a$_{3}$m$_{s}$ values reduce (second panel of figure~\ref{FFF}). In addition, the threshold density values get closer by decrease of strangeness in a way that the nucleon-nucleon interactions do not have any effect on this quantity. However, AV6'+TNI potential results in higher kaon fraction compared to other potentials in both strangeness values.} Moreover, one can observe that the kaon-nucleon coupling parameter, a$_{3}$m$_{s}$=-222 MeV, results in the highest values for proton abundance in all potentials. On the other hand, it can be seen that the proton fraction behaves as the increasing function of density for the values of a$_{3}$m$_{s}=-178$ and -134 MeV, while a$_{3}$m$_{s}=-222$ MeV leads to the slight reduction in the value of proton fraction at higher densities.

In figure~\ref{F2}, the candidate EoSs for four values of strangeness, i.e, a$_{3}$m$_{s}$=-310, -222, -178, and -134 MeV are displayed. Note that the pressure and thus the equation of state of neutron star matter can be obtained through the thermodynamic relation
{\begin{eqnarray}\label{tener}
           P=\rho\frac{\partial}{\partial \rho}{\varepsilon}-\varepsilon.
 \end{eqnarray}}
Furthermore, the results obtained for the EoSs without kaon condensation, which have been discussed in our previous works \cite{Asadi2018, sharifi}, are presented by solid black lines in this figure. It can be noticed that in this figure, the kaon condensation with all values of a$_{3}$m$_{s}$ results in the reduction of pressure, in particular for the densities close to the threshold density and also lower amounts of a$_{3}$m$_{s}$. On the other hand, the softening of the EoSs takes place in the whole range of density for all values of strangeness except a$_{3}$m$_{s}$=-310 MeV for the AV6$'$+TNI case, which stiffens the EoS at the end of the density range.

Figure~\ref{F3} represents the results of figure~\ref{F2} applying the Maxwell construction for two values of strangeness, -222 and -178 MeV, in order to resolve the unstable region of EoS. As it is indicated by figure~\ref{F2}, the condensed EoS with the strangeness of -310 MeV is too soft, especially in the low-density region that it is not possible to employ the Maxwell construction. Note that kaons appear at higher amounts of density as the value of a$_{3}$m$_{s}$ shifts from -222 MeV to -178 MeV for all three EoS models. It is worth mentioning that the range of density with the constant pressure resulting from employing the Maxwell construction is broader for a$_{3}$m$_{s}$=-222 MeV in comparison with a$_{3}$m$_{s}$=-178 MeV in all models. However, AV8$'$+TNI as the softest EoS has the widest density range with constant pressure. Moreover, one can observe that the value of a$_{3}$m$_{s}$=-222 MeV stiffens the condensed EoSs at very high density values compared to the other value of a$_{3}$m$_{s}$ in this figure.

{\section{Mass-Radius relation}}
We plot the relation between mass and radius of neutron stars for all three candidate EoS models with and without kaon condensation in figure~\ref{F4}, which is obtained by solving the Tolman-Oppenheimer-Volkoff (TOV) equations~\cite{Tolman,Oppenheimer}
\begin{eqnarray}
&&\frac{dP}{dr}=-\frac{\epsilon m}{r^2}(1+\frac{P}{\epsilon })(1+\frac{4\pi Pr^3}{m})(1-\frac{2m}{r})^{-1}\nonumber\\&&
\frac{dm}{dr}=4\pi r^2 \epsilon.
  \end{eqnarray}
Here $\varepsilon$ and $p$ correspond to the energy density and pressure, respectively. It should be pointed out that a$_{3}$m$_{s}$=-310 MeV is excluded from our results of structural properties due to the intense softness that imposes on the relevant EoS. The mass-radius relation of EoSs without koan condensation, which is shown by the solid blue line, indicates that they all lead to the maximum mass with values greater than 2 M$_\odot$. However, one can notice that the shift of a$_{3}$m$_{s}$ value from -134 to -222 MeV affects the maximum mass of neutron stars such that it decreases with the softness of the condensed EoSs. In other words, there is a considerable difference in the maximum mass of neutron stars with smaller values of strangeness, i.e -178 and -222 MeV, compared to those without kaon condensation. Some observational constraints are presented in figure~\ref{F4}, e.g. the maximum mass of non-rotating NSs: 2.01$_{-0.04}^{+0.04}\leq M_{TOV}/M_{\odot}\lesssim 2.16_{-0.15}^{+0.17}$ \cite{Rezzollaa}, the radius of stellar structure of 1.6 M$_{\odot}$ non-rotating NSs: $R_{1.6}>$10.68$_{-0.04}^{+0.15}$ km \cite{Bauswein}, the maximum radius of a 1.4 M$_{\odot}$ NS: 13.6 km \cite{Annala}, a lower boundary on the maximum mass of non-rotating NSs: M$_{TOV}>2.08^{+0.04}_{-0.04}M_{\odot}$ \cite{Mostt}, {and the radius of 1.4 M$_{\odot}$ NS: 12.33$_{-0.81}^{+0.76}$ km (PP model) and 12.18$_{-0.79}^{+0.56}$ km (CS model) \cite{Raaijmakers}}. It is apparent from the figure that the results obtained by a$_{3}$m$_{s}$=-134 MeV agree well with the maximum mass constraint (GW170817 excluded: Rezzolla \textit{et al.} \cite{Rezzollaa}) and the lower limit on the maximum mass of non-rotating NSs \cite{Mostt} for the EoSs with AV18+TNI and AV6'+TNI. However, the other values of a$_{3}$m$_{s}$ do not lead to consistent results with the mentioned constraints considering all the EoS models. The only condensed EoS that cannot satisfy the R$_{1.6}$ constraint \cite{Bauswein} is the one with a$_{3}$m$_{s}$=-222 MeV for all models. It is to be noted that all the EoS models considered in this figure fulfill the right-hand band, (GW170817 excluded: Annala \textit{et al.} \cite{Annala}). {Furthermore, one can observe that there is an unstable region after the appearance of kaons for a$_{3}$m$_{s}$=-222 MeV despite other values, which manifests the existence of NSs with the same mass but different radius (twin stars). It refers to the various composition of the NS core: pure hadronic matter and kaon condensed matter. {It is also clear that this value of strangeness does not fulfill radius constraints for the 1.4 M$_{\odot}$ NS reported by NICER's mass-radius estimate of PSR J0740+6620 and multimessenger observations
considering all the EoS models \cite{Raaijmakers}.}}

Table~II displays the obtained values for the maximum mass, the corresponding {central density,} the radius, and the compactness of neutron stars {(also radius and dimensionless tidal deformability of 1.4 M$_{\odot}$ NS)} considering kaon condensation with the three discussed values of a$_{3}$m$_{s}$. The values for the EoSs without kaon condensation are also presented in the second column of this table. As it is obvious, the decrease of a$_{3}$m$_{s}$ leads to the reduction in the value of the maximum mass and the corresponding radius for all the models. For instance, the value of the maximum mass (radius) in the case of AV18+TNI changes from 2.05 M$_{\odot}$ (10.52 km) to 1.86 M$_{\odot}$} (8.40 km) considering a$_{3}$m$_{s}$=-134 and -222 MeV, respectively. However, the compactness $\beta$ increases with the reduction of the strangeness, in particular for the case with a$_{3}$m$_{s}$=-222 MeV which has the value of 0.332 for both AV6'+TNI and AV8'+TNI.
\begin{figure}[!ht]
\centerline{\includegraphics[scale=0.7]{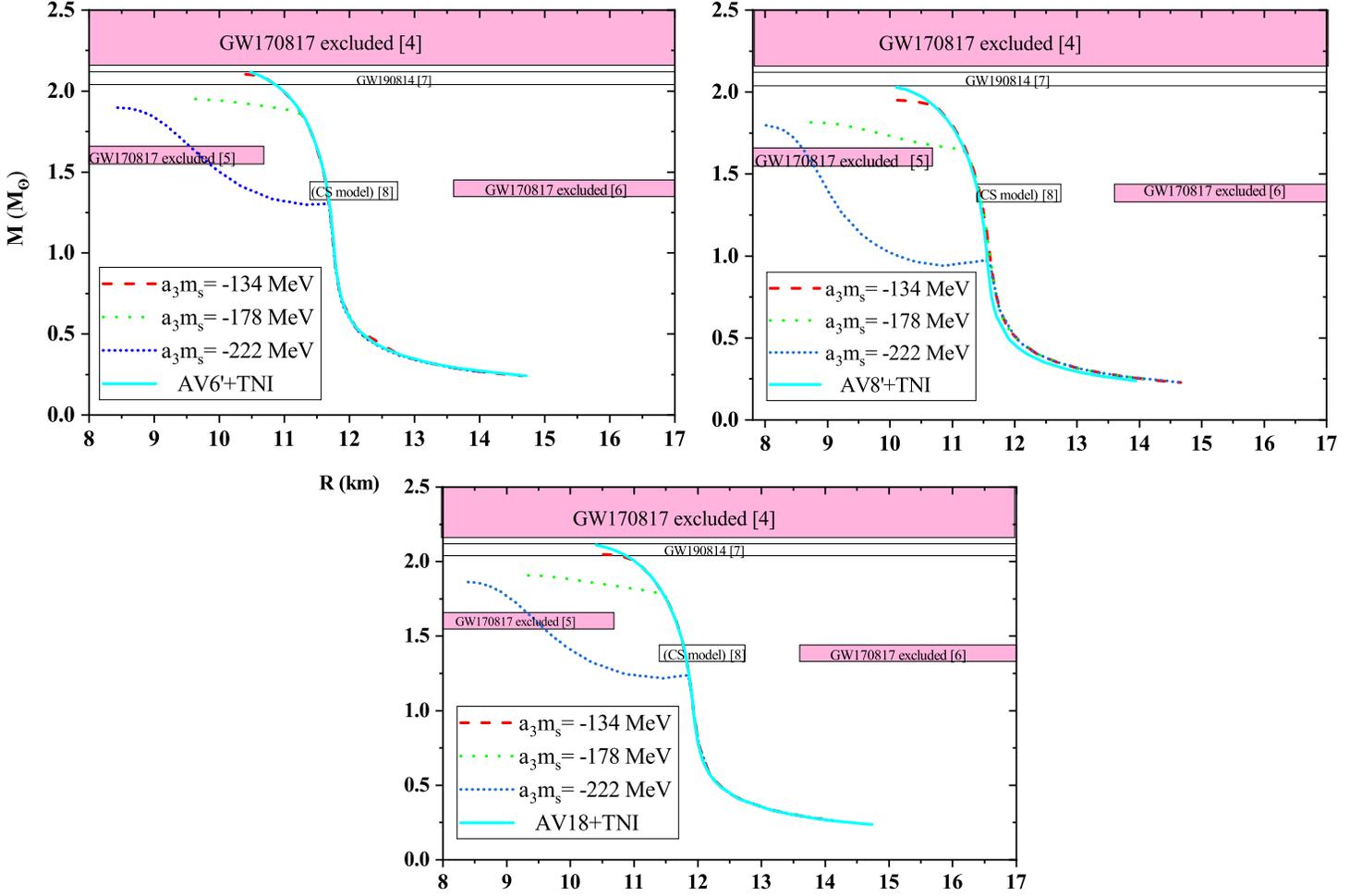}}
\caption{\small {Mass-radius relation for the candidate EoSs discussed in the text.}}\label{F4}
\end{figure}
\begin{table*}
 {\caption{{\label{1} Maximum mass along with corresponding {central density,} radius, and compactness of neutron stars with and without the kaon condensation. {Moreover, the values of the radius and dimensionless tidal deformability related to the 1.4 M$_{\odot}$ NS structure are also presented in this table.}}}}

\begin{tabular}{ccccc}
  \hline
   \hline
AV6$'$+TNI \ \ \ &w/o kaon \ \ \ & a$_3$m$_{s}$=-134 MeV \ \ \ &a$_3$m$_{s}$=-178 MeV \ \ \ &a$_3$m$_{s}$=-222 MeV \\
 \hline
   M$_{max}$ (M$_{\odot}$)&2.12&2.10 &1.95&1.90\\
  \hline
  {$\rho_{c}$ ($fm^{-3}$)}&{1}&{1.1}&{1.3}&{1.6}\\
\hline
  R$_{max}$ (km)&10.48&10.40&9.63&8.43\\
\hline
   $\beta$&0.298&0.299&0.299&0.332
\\
   \hline
   {R$_{1.4}$ (km)}&{11.65}&{11.65}&{11.65}&{10.40}
\\
   \hline
 {$\Lambda_{1.4}$}&{292}&{292}&{292}&{103}
\\
   \hline
  AV8$'$+TNI
  \\
  \hline
  \hline
  M$_{max}$ (M$_{\odot}$)\ \ \ &2.03&1.95&1.81&1.80\\
  \hline
  {$\rho_{c}$ ($fm^{-3}$)}&{1.1}&{1.2}&{1.6}&{1.7}\\
\hline
  R$_{max}$ (km)&10.13&10.12&8.73&8.01
  \\
   \hline
   $\beta$&0.296&0.285&0.307&0.332\\
   \hline
   {R$_{1.4}$ (km)}&{11.40}&{11.40}&{11.40}&{9}
\\
   \hline
   {$\Lambda_{1.4}$}& {255}&{255}& {255}& {35}
\\
\hline
   \hline
  AV18+TNI
  \\
  \hline
  M$_{max}$ (M$_{\odot}$)\ \ \ &2.11&2.05&1.91&1.86\\
  \hline
 {$\rho_{c}$ ($fm^{-3}$)}&{1.05}&{1.1}&{1.4}&{1.6}\\
\hline
  R$_{max}$ (km)&10.40&10.52&9.34&8.40
  \\
  \hline
  $\beta$&0.299&0.288&0.301&0.328\\
  \hline
 {R$_{1.4}$ (km)}&{11.79}&{11.79}&{11.79}&{10.02}\\
  \hline
   {$\Lambda_{1.4}$}& {314}& {314}&{314}& {74}
\\
   \hline
   \hline
  \end{tabular}
  \end{table*}
\begin{figure}[!ht]
\centerline{\includegraphics[scale=0.7]{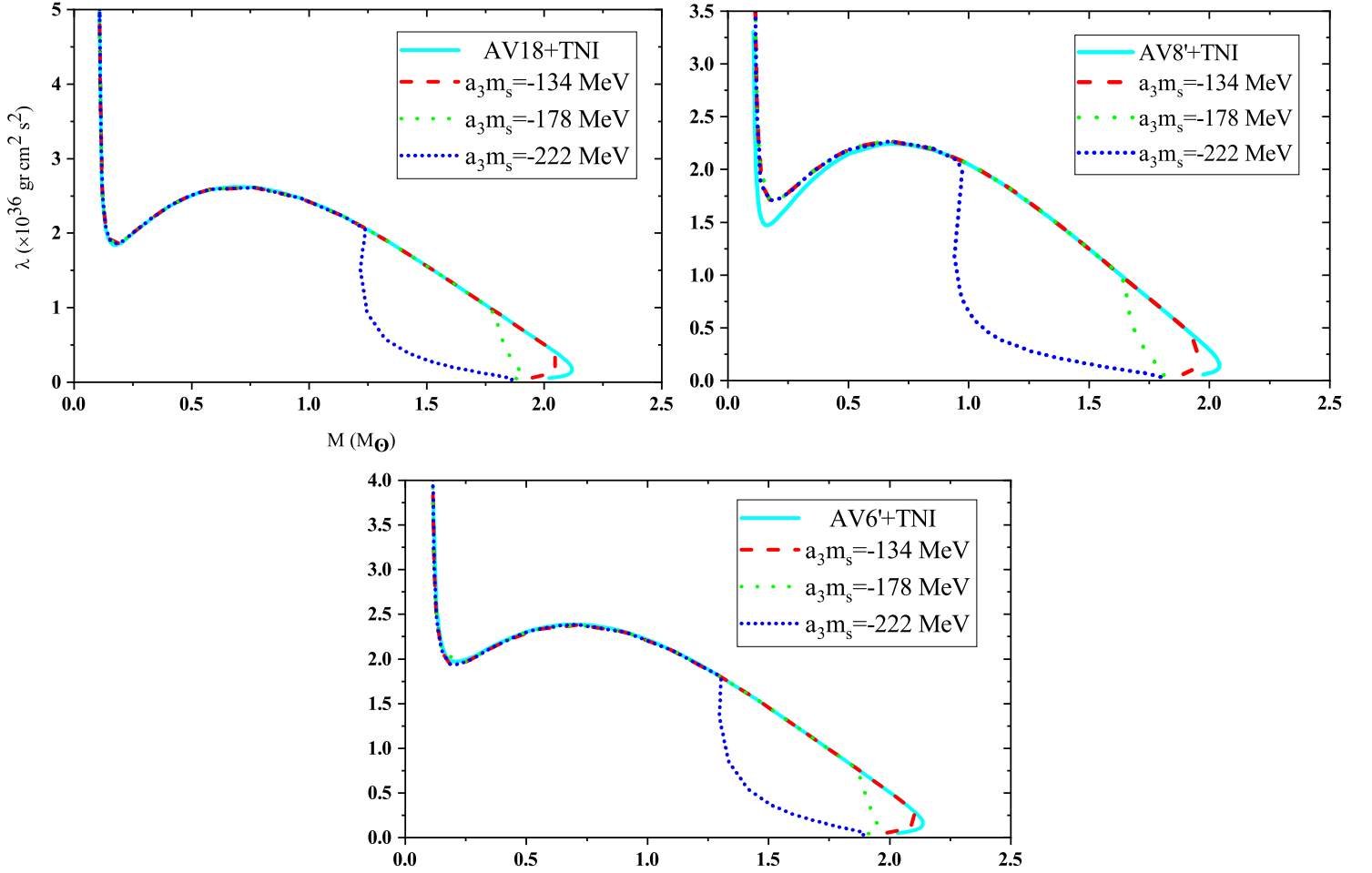}}
\caption{\small {Tidal deformability versus neutron star mass for the candidate EoSs with selected values of strangeness.}}\label{F5}
\end{figure}
\begin{figure}[!ht]
\centerline{\includegraphics[scale=0.6]{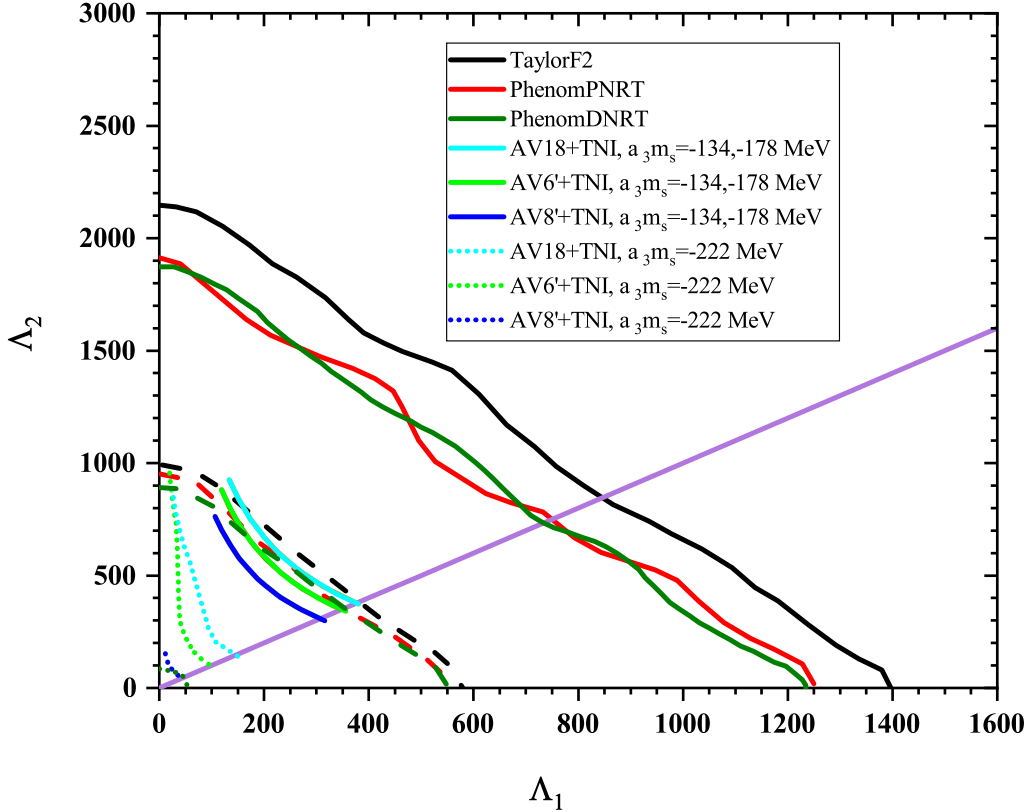}}
{\caption{\small {Dimensionless tidal deformability parameters of a binary neutron star considering the candidate EoSs with the selected values of strangeness. It is to be noted that the values of a$_{3}$m$_{s}$=-134 and -178 MeV lead to similar results. The 50\% (dashed lines) and 90\% (solid lines) credible intervals are represented for three waveform models, i.e. TaylorF2, PhenomPNRT, and PhenomDNRT.}}}\label{F6}
\end{figure}
\begin{figure}[!ht]
\centerline{\includegraphics[scale=0.6]{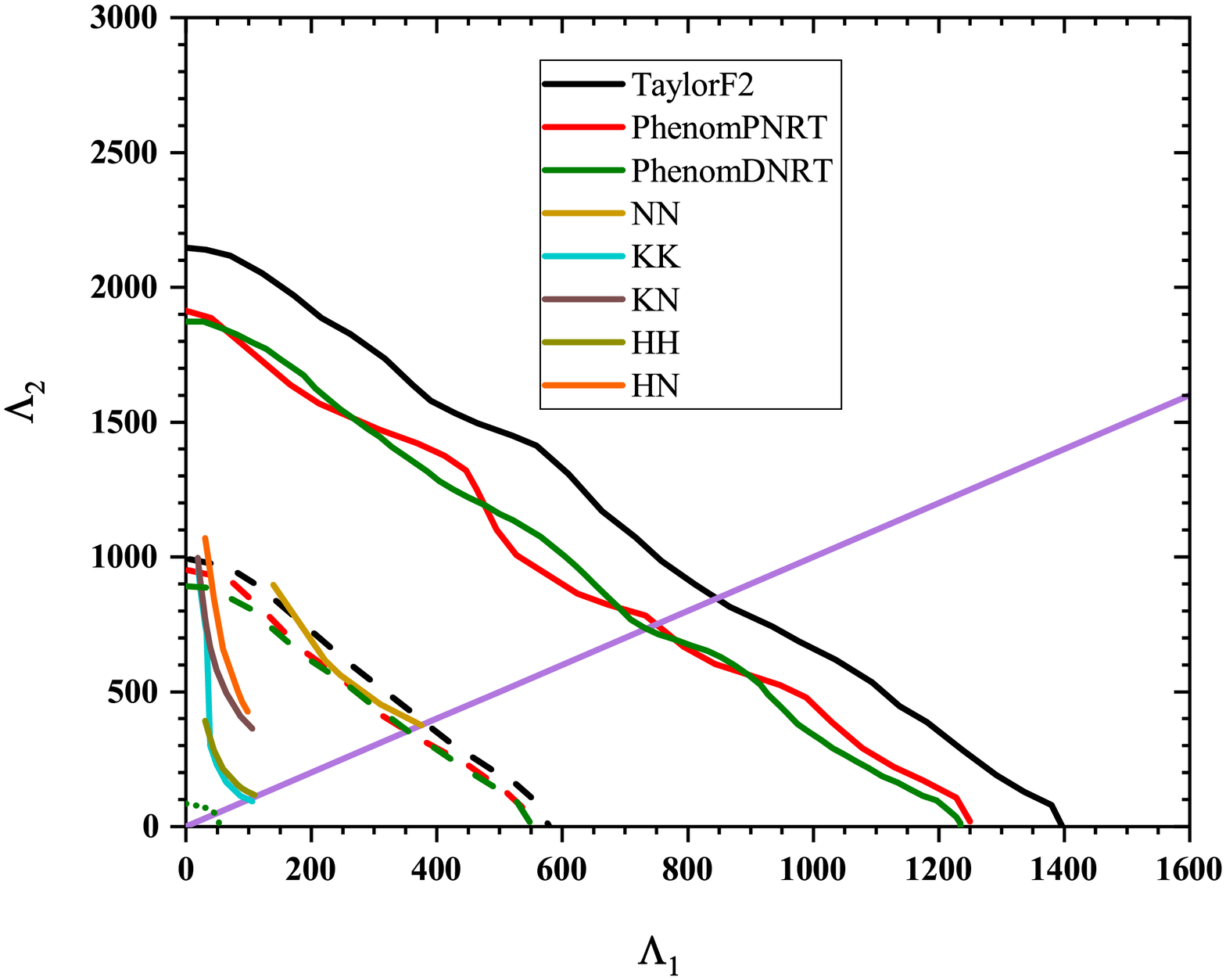}}
{\caption{\small {Same as in Figure 7, but for three possible components of a binary system considering kaon condensation with the model of AV18+TNI potential and a$_{3}$m$_{s}$=-222 MeV. For comparison, our results for the first-order phase transition to quark matter with the same EoS are also presented here \cite{sharifi1}.}}}\label{F7}
\end{figure}

\section{Tidal deformability}
{GW170817 is an observed merger event that has set some constraints on the tidal deformability of compact stars $\lambda$, which is related to their EoS through the radius as $\lambda=\frac{2}{3}k_{2}R^{5}$, where $k_{2}$ is the tidal love number corresponding to l=2 \cite{Hinderer1, Hinderer2}. It is intriguing to study the behaviour of this quantity with respect to the observable like the mass of NSs for which the kaon condensation is considered together with our EoSs in figure \ref{F5}. As it is obvious, the behaviour of $\lambda$ as a function of mass with the value of a$_{3}$m$_{s}$=-134 MeV {deviates slightly from those without kaons at the maximum value of the neutron star mass for all the models. One can notice that as strangeness decreases, this deviation becomes considerable, particularly for a$_{3}$m$_{s}$=-222 MeV in which the figure shifts to the lower values of mass in the presence of kaons. Moreover, as the radius decreases with the softening of the EoS, the tidal deformability undergoes a reduction in values due to the mentioned dependence.}}

{The dimensionless tidal deformability, which can be obtained by $\Lambda=\frac{2}{3}k_{2}\beta^{5}$ with $\beta$ being the compactness of neutron star, is the best parameter to be compared with the limits deduced from the observation. Therefore, figure 7 is displayed in the form of $\Lambda_{1}-\Lambda_{2}$ plot considering a binary system with the component masses of $m_{1}$ (higher value) and $m_{2}$ (lower value) for the same selection of EoSs in the previous figures. It should be pointed out that the binary system is established with the chirp mass $\mathcal{M}=(m_{1}m_{2})^{3/5}/(m_{1}+m_{2})^{1/5}$ fixed at the measured value of 1.186 M${\odot}$ \cite{abbot2019}. Moreover, this figure involves the 50\% (dashed black, red, and dark green lines) and 90\% (solid black, red, and dark green lines) of credible intervals related to the TaylorF2, PhenomPNRT, and PhenomDNRT waveform models for the low spin case \cite {abbot2019}. It is worth mentioning that the results obtained for the values of a$_{3}$m$_{s}$=-134 and -178 MeV are exactly the same in which one cannot distinguish these two lines from each other in all cases of EoSs. One can therefore observe that considering these two values of strangeness, the EoS with AV18+TNI results in larger values of tidal deformability which lie between the 50\% upper limits of the mentioned waveform models, while AV6$'$+TNI is approximately near this region and AV8$'$+TNI shifts to the lower values of tidal deformability as the softest EoS. The short-dotted cyan, green, and blue lines correspond to the results obtained by the strangeness of -222 MeV with selected EoSs, which lie below the 50\% upper limits due to the softness of EoSs. As it is expected, this value of strangeness intensifies the softness of AV8$'$+TNI such that it includes the smallest values of tidal deformability compared to the other models. It means that the neutron star described by the EoS with AV8$'$+TNI and a$_{3}$m$_{s}$=-222 MeV is less deformed due to its higher values of compactness.

{In order to take account of the binary system in which the components have the same mass but different size, we demonstrate the $\Lambda_{1}-\Lambda_{2}$ plot for the EoS of AV18+TNI together with a$_{3}$m$_{s}$=-222 MeV in figure 8. For comparison, we also present our previously obtained results produced by AV18+TNI in which the first-order phase transition to the quark matter has been considered \cite{sharifi1}. It is obvious from this figure that Kaon star-Neutron star (KN) combination lies in a similar region to that of Hybrid star-Neutron star (HN), which are mostly below the 50\% upper limits. The results obtained for the Kaon star-Kaon star (KK) combination are consistent with those of Hybrid star-Hybrid star (HH) considering the larger amounts of mass ratio, $q=m_{2}/m_{1}$, both of which shift to the region with smallest values of tidal deformability. Thus, one can conclude that the phase transition from hadronic matter to the kaon condensation matter has no considerable impact on the values of tidal deformability compared to the case of quark matter within our selected EoS. Although the KK and KN combinations with a$_{3}$m$_{s}$=-222 MeV are allowed in this figure, they have been ruled out by the constraint presented by Bauswein \textit{et al.} \cite{Bauswein} in the mass-radius diagram.}

\section{Summary and Conclusions}
In this paper, we have calculated the equation of state of kaon condensed neutron star matter applying the LOCV approach. We have taken into account the two-body potentials such as AV6$'$, AV8$'$, and AV18 with the three nucleon interaction included. It is worth mentioning that for the kaon-nucleon interaction we have used the chiral model, which was first introduced by Kaplan and Nelson. Furthermore, we have computed the structural properties and also the tidal parameters of neutron stars considering kaon condensation and checked the validity of our results with the recent observational limits. Considering the kaon condensed EoS model with AV18+TNI and AV6$'$+TNI together with a$_3$m$_s$=-134 MeV, all the new observational constraints on the maximum mass and the radius of NSs are fulfilled as well as dimensionless tidal deformability shown in the form of $\Lambda_{1}-\Lambda_{2}$ plot. It is found that the reduction in the value of strangeness lowers the maximum mass of neutron star in a way that the results obtained by a$_3$m$_s$=-222 MeV lead to the structural properties that are not consistent with recent observational results. It is to be noted that the structure of NS undergoes instability after kaon condensation with this value of a$_3$m$_s$. Therefore, in order to consider the binary system with the same mass but different sizes, we have used the EoS model with AV18+TNI and a$_3$m$_s$=-222 MeV. We have compared our results with the case in which the quark matter phase is present at the core of neutron star. It turned out that there is no considerable influence on the values of tidal deformability in comparison to the case with the first-order phase transition to the quark matter.

\section*{Acknowledgments}
D.A-C. acknowledges support from the Bogoliubov-Infeld program for collaboration between JINR and Polish
Institutions and to the COST Action CA16214 "PHAROS" for networking activities.



\end{document}